\documentclass[sigconf]{acmart}  

\pdfoutput=1

\usepackage{booktabs} 
\usepackage[linesnumbered,ruled,vlined]{algorithm2e}
\usepackage{setspace}
\usepackage{array}

\SetKw{Continue}{continue}

\usepackage{tikz}
\usetikzlibrary{intersections}
\usetikzlibrary{patterns}
\usetikzlibrary{shapes,backgrounds}
\usepackage{adjustbox}
\usepackage{fixltx2e}
\usetikzlibrary{shapes.geometric}
\usepackage{pgfplots}
\usetikzlibrary{arrows}

\usepackage{caption, subcaption}

\usepackage{amsthm}
\usepackage{amsmath}
\theoremstyle{definition}
\newtheorem*{probdef}{\textbf{Problem Definition}}

\newtheorem{thm}{Theorem}
\newtheorem{corollary}{Corollary}

\renewcommand\footnotetextcopyrightpermission[1]{} 
\thanks{An abridged version of this manuscript has been accepted for publication as a short paper at ACM SIGSPATIAL 2018.}

\begin{document}
\title{In-Route Task Selection in Crowdsourcing}

\author{Camila F. Costa}
\affiliation{%
  \institution{University of Alberta, Canada}
  }
\email{camila.costa@ualberta.ca}

\author{Mario A. Nascimento}
\affiliation{%
  \institution{University of Alberta, Canada}
  }
\email{mario.nascimento@ualberta.ca}

\begin{abstract}
One important problem in crowdsourcing is that of assigning tasks to workers. We consider a scenario where a worker is traveling on a preferred/typical path (e.g., from school to home) and there is a set of tasks available to be performed. Furthermore, we assume that: each task yields a positive reward,  the worker has the skills necessary to perform all available tasks and  he/she is willing to possibly deviate from his/her preferred path as long as he/she travels at most a total given distance/time.  We call this problem the {\em In-Route Task Selection} (IRTS) problem and investigate it using the skyline paradigm in order to obtain the exact set of non-dominated solutions, i.e., good and diverse solutions yielding different combinations of smaller or larger rewards while traveling more or less. This is a practically relevant problem as it empowers the worker as he/she can decide, in real time, which tasks suit his/her needs and/or availability better. After showing that the IRTS problem is NP-hard, we propose an exact (but expensive) solution and a few others practical heuristic solutions.  While the exact solution is suitable only for reasonably small IRTS instances, the heuristic solutions can produce solutions with good values of precision and recall for problems of realistic sizes within practical, in fact most often sub-second, query processing time.
\end{abstract}


\maketitle

\section{Introduction}
\label{sec:intro}

Crowdsourcing is a relatively new computing paradigm which relies on the contributions of a large number of workers to accomplish tasks, such as image tagging and language translation. The increasing popularity of mobile computing led to a shift from traditional web-based crowdsourcing to spatial crowdsourcing \cite{tong2017spatial}. Spatial crowdsourcing consists of location-specific tasks, submitted by requesters, that require people to physically be at specific locations to complete them. Examples of these tasks include taking pictures or answering questions about a certain location in real time. Tasks are assigned to suitable workers based on a particular objective, such as maximizing the number of assigned tasks, maximizing a given matching score, minimizing the total amount of reward paid out by task requesters or the total reward earned by workers after deducting traveling costs. 
(We defer the discussions about those approaches when presenting the related work in Section~\ref{sec:related}.)

Traditionally,
in the cases where a worker is assigned to multiple tasks, the travel cost between tasks is not typically taken into account. However, that cost directly affects the number of tasks the worker will be able to perform. Thus, even if those tasks are spatially close from the worker, they may not be completed depending, for example, on the worker's time/distance budget.
Therefore, in this paper we consider the  more generic problem of finding a {\em task schedule} for a worker, i.e., a feasible sequence of tasks, a problem that has also been considered in \cite{deng2013maximizing,deng2015task}.

However, differently from any  work that we are aware of, we consider a scenario where a worker is (or will be) traveling along a predetermined path and he/she is willing to possibly perform tasks while on that path. For instance, consider a user and his/her preferred path, say a particular bicycle path or bus route from school to home. On the one hand, it would make sense to consider performing tasks that minimize the detour from the preferred path. On the other hand, considering that each task is associated with a reward, it would also make sense to maximize the total reward received for performing tasks.  We refer to this problem as \emph{In-Route Task Selection} (IRTS).  The added novelty and non-trivial complexity of this problem comes from the fact that there are two \emph{competing criteria} to be optimized at the same time: deviation and reward.  


More formally, the IRTS problem can be formulated as follows.  Given a preferred path $P^*$, a budget $b$ and a set of geographically located tasks $T$, the IRTS problem aims at maximizing the total reward received by the worker while minimizing the total detour from $P^*$ incurred for traveling to the location of tasks, considering that the total traveling cost does not exceed $b$.  

\begin{figure}[htb!]
\centering
    \includegraphics[width=0.6\linewidth]{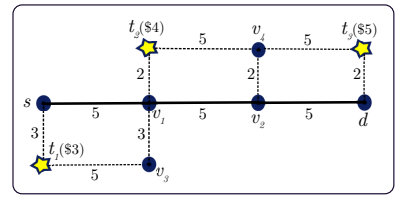}
    \caption{Preferred path $P^*$ and tasks $t_1$, $t_2$ and $t_3$ with their corresponding rewards. 
    }
    \label{fig:irts}
\end{figure}

In order to illustrate IRTS, consider the simple scenario shown in Figure~\ref{fig:irts}. The path in bold represents the worker's preferred path $P^*= \langle s, v_1, v_2, d \rangle$, say from school to home. Moreover, there are three available tasks $T = \{t_1, t_2, t_3\}$ associated with their corresponding rewards, \$3, \$4 and \$5, respectively. Also, assume that $b = 21$. On the one hand, if the worker wants to minimize the detour from $P^*$ incurred from traveling to a task location, a possible path is $P^1= \langle s, v_1, t_2, v_1, v_2, d \rangle$ (with reward \$4 and detour 4). On the other hand, if the user wants to maximize his/her reward, the path $P^2= \langle s, v_1, t_2, v_4, t_3, d \rangle$ (with reward \$9 and detour 14) would be the best option. Let us now consider other alternative paths. 
$P^3= \langle s, t_1, s, v_1, v_2, d \rangle$ yields a total detour of 6 and reward of \$3, $P^4= \langle s, v_1, v_2, d, t_3, d \rangle$ yields a total detour of 4 and reward of \$5, while $P^5= \langle s, v_1, t_2, v_1, v_2, d, t_3, d \rangle$ yields a total detour of 8 and reward of \$9. However, since the total cost of $P^5$ is 23, which is greater than the specified budget, $P^5$ is not a feasible option.

As shown above, a single route does not typically optimizes both criteria, i.e., minimize detour and maximize reward, at the same time. A simplistic way to address this problem is to combine both criteria  and optimize the combined value. However, finding a single meaningful function means weighting the importance of each criterion. This would depend primarily on the worker's preferences, which may be not obvious or clear from the outset, adding an extra parameter to the problem at query time.  Moreover, we believe that it is of practical relevance to empower the worker him/herself to consider all interesting alternatives.

\begin{figure}[htb!]
\centering
    \includegraphics[width=0.4\linewidth]{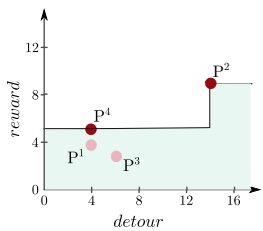}
    \caption{Skyline for the example illustrated in Figure~\ref{fig:irts} and Table~\ref{tab:paths}.     }
    \label{fig:skyline}
\end{figure}

Fortunately, a more principled way to deal with the IRTS problem is to determine all results that are optimal under any arbitrary combination of the two criteria.  That can be achieved by using the notion of skyline queries \cite{borzsony2001skyline}.  In generic terms, the result set of a skyline query contains objects which are not dominated by any other one. An object $o_i$ is dominated by another object $o_j$ if, for each criterion, $o_i$ is at most as good as $o_j$, and, for at least one criterion, $o_j$ is strictly better than $o_i$. For instance, consider the paths (along with their detours and rewards) shown in Table~\ref{tab:paths} and based on Figure~\ref{fig:irts}. Path $P^1$ is dominated by path $P^4$, since both of them yield the same detour, but $P^4$ has a higher reward. Similarly, $P^3$ is also dominated by $P^4$ since $P^4$ is better than $P^3$ in both criteria. On the other hand, paths $P^2$ and $P^4$ are non-dominated, since none of them is better than the other both is terms of detour \textit{and} reward. Therefore, $P^2$ and $P^4$ are equally interesting and should be offered as alternatives to the user, who can decide by him/herself how prioritize the trade-off between deviation and the reward. Figure~\ref{fig:skyline} illustrates the concept of skyline queries. It shows the corresponding detours and rewards of these four candidate paths. The linked dots denote the non-dominated paths and the shaded area contains the dominated ones.

\begin{small}
\begin{table}[!htb]
\centering
\caption{Paths extracted from Figure~\ref{fig:irts} with their corresponding rewards and detours, and whether they are dominated and by whom.} 
\label{tab:paths}
\begin{tabular}{|c|c|c|c|}
\hline
\textbf{Path} & \multicolumn{1}{l|}{\textbf{Reward}} & \multicolumn{1}{l|}{\textbf{Detour}} & \multicolumn{1}{l|}{\textbf{Dominated by}} \\ \hline
\hline
$P^1$         & 4                                    & 4 & $P^4$                                   \\ \hline
$P^2$         & 9                                    & 14 & non-dominated                                   \\ \hline
$P^3$         & 3                                    & 6 & $P^4$                                   \\ \hline
$P^4$         & 5                                     & 4 & non-dominated                                  \\ \hline
\end{tabular}
\end{table}
\end{small}

The main contributions we offer in this paper are two-fold.  We present the IRTS problem, a new extension of the crowdsourcing problem in spatial data, which empowers the worker with informed choices regarding the available tasks he/she can choose from. After showing that the problem is NP-hard, we present few practically alternative solutions that can be used to solve instances of the IRTS problem in a city scale efficiently.

The remainder of this paper is structured as follows. In Section~\ref{sec:related} we present relevant related work and contrast it to ours. We present the formal definition of the  IRTS problem, showing it is NP-hard, in Section \ref{sec:problem_definition}. Our proposed solutions are presented in Section~\ref{sec:solutions}, followed by their experimental evaluations using real datasets in Section~\ref{sec:experiments}. Finally, Section~\ref{sec:conclusion} presents a summary of our findings and  suggestions for future work.
\section{Related Work}
\label{sec:related}

The literature in spatial crowdsourcing presents many different ways to assign tasks to workers, for instance,  maximizing the number of assigned tasks, 
\cite{kazemi2012geocrowd, to2016real}, 
maximizing a given matching score \cite{tong2016online, song2017trichromatic, zheng2016mutual,cheng2015reliable}
or minimizing the total amount of reward paid out by requesters, while maximizing the number of assignments \cite{dang2013towards}.


Kazemi and Sahabi \cite{kazemi2012geocrowd} study the maximum task assignment (MTA) problem in spatial crowdsourcing, which aims at maximizing the overall number of assigned tasks and considers that a worker only accepts tasks within his/her spatial region and is only willing to perform up to a predetermined number of tasks. To et al. \cite{to2016real} introduce a framework for crowdsourcing hyper-local information. A task can only be answered by workers who are already within a radius $r$ from the task location at a time when the task is valid. The goal is to maximize task assignment given a budget, which is the maximum number of workers that can be selected. 

Tong et al. \cite{tong2016online} study the Global Online Micro-task Allocation in spatial crowdsourcing (GOMA) problem, which  aims at maximizing a total matching utility. 
The utility for a task-worker pair $(t,w)$ is given by the payoff of task $t$ times the success ratio of $w$ completing tasks. 
Song et al. \cite{song2017trichromatic} propose the trichromatic online matching (TOM) in real-time spatial crowdsourcing  problem, which aims at matching three types of objects, namely tasks, workers and workplaces. The goal is to maximize a total utility score, which represents the satisfaction of the matching involving the corresponding task requester, worker
and workplace. 
Zheng and Chen \cite{zheng2016mutual} study the task assignment with mutual benefit awareness (TAMBA) problem. TAMBA aims at maximizing the mutual benefit of the workers and tasks, which is measured in terms of the expected answer quality for a worker $w$ and task $t$, given by the probability of $w$'s acceptance of $t$ multiplied by the expected rating of $w$'s completing $t$. 

Within the context of reward-based task assignment, Dang and Cao \cite{dang2013towards} propose the Maximum Task Minimum Cost Assignment (MTMCA) problem. MTMCA considers that workers have multiple skills and that each task has a type. The goal is to maximize the number of assignments and subsequently minimize the total amount of money spent by the requesters, assuming that the price of tasks is attributed by workers. The Multi-Skill Spatial Crowdsourcing (MS-SC) problem is presented in \cite{cheng2016task}. It aims at assigning multi-skilled workers to complex spatial tasks such that skills between workers and tasks match with each other, and workers benefits are maximized. Each task $t_j$ is associated with a budget $B_j$, which represents the maximum amount the requester is willing to pay for $t_j$. The workers rewards is given by the sum of the remaining budget of the completed tasks after deducting traveling costs. 

Differently from the works above, \cite{deng2013maximizing,deng2015task} deal with assigning a sequence of tasks to workers and, thus, take the travel cost between tasks into account, as we do.
In \cite{deng2013maximizing}, the authors focus on maximizing the number of tasks performed by a single worker assuming that the worker is willing to perform up to a predefined maximum number of tasks and that he/she must arrive at the task location before its deadline. An extension of that problem is proposed in \cite{deng2015task}, which aims to maximize the overall number of tasks performed by multiple workers considering that each task is assigned to at most one worker.
As in \cite{deng2013maximizing}, we focus on the scenario where workers self-select the tasks they want to perform from a list of published tasks, termed Worker Selected Tasks (WST) \cite{kazemi2012geocrowd} mode. However, differently from \cite{deng2013maximizing,deng2015task}, we assume that a worker is willing to perform tasks while traveling along a predetermined path but likely without deviating too much from it.  Additionally, we consider that each task is associated with a reward and that the user also wants to maximize the total reward received for performing tasks. 

All of the works above focus on either optimizing a single criterion or on also optimizing a secondary criterion, which serve as a tie-breaker. On the other hand, Cheng et al. \cite{cheng2015reliable} study the reliable diversity-based spatial crowdsourcing (RDB-SC) problem, which assigns workers to tasks such that tasks can be accomplished with high reliability and spatial/temporal diversity. Thus, there are two criteria to be optimized simultaneously: reliability and diversity. Although they do not find a skyline set, they rely on the notion of dominance and  select the \emph{one} solution that dominates the most solutions as the best one. 
Unlike in \cite{cheng2015reliable}, we provide the worker with choices by returning the optimal skyline set of solutions.


The IRTS problem can be also seen as an interesting combination of two seemingly unrelated problems: In-Route Nearest Neighbor queries and the Orienteering Problem. 

The problem of searching for nearest neighbors with respect to a given (preferred) path has been previously defined as In-Route Nearest-Neighbor (IRNN) queries ~\cite{shekhar2003processing}. Within the context of multiple competing criteria in IRNN queries, \cite{AhmadiCN17} focuses on the trade-off yielded by minimizing the detour incurred for visiting a \emph {single} point of interest (POI) and also minimizing the total cost of the path at the same time. On the other hand, \cite{huang2004route} aims at minimizing the cost for reaching a POI, as opposed to the total cost of the path, and the detour incurred. 
In the Orienteering Problem (OP) \cite{golden1987orienteering}, it is given a graph $G(V,E)$ where each vertex $v \in V$ is associated with a positive score, and a budget $b$. OP aims at finding the route from a given starting point $s$ that maximizes the total score while the total travel cost does not exceed $b$.

The main differences between IRTS and the IRNN and OP are the following. IRNN considers deviating towards one single POI, while IRTS considers multiple tasks (which are akin to POIs in the IRNN context).  In fact, IRTS can be seen as a generalization of the IRNN problem.  The OP problem does not consider the notion of trade-off between travel cost and rewards at all, and in this respect IRTS can be considered a non-trivial extension to OP.
Also, OP considers only the starting point of the traveler (worker),  and thus there is no concept of ``destination'', whereas IRTS considers the worker's preferred path, i.e., we contextualize the IRTS problem w.r.t. the worker's preferences/plans.


\section{Preliminaries}
\label{sec:problem_definition}

We assume that the worker's movement is constrained by an underlying road network, which is modeled as an undirected graph $G(V,E,C)$, where $V$ is a set of vertices that represent the road intersections and end-points, $E$ is the set of edges containing all road segments and $C$ indicates the costs of edges in $E$. In our case, the cost of an edge connecting vertices $v_i$ and $v_j$ is given by the length of the road network segment that connects those vertices and is denoted by $c(v_i,v_j)$. 

We define a path $P^i=\langle  v^i_{1},v^i_{2},...,v^i_{n} \rangle$ in $G$ as a sequence of vertices such that any two consecutive vertices $v^i_{j}$ and $v^i_{j+1}$, for $1 \leq j < n$, are directly connected by an edge $(v^i_{j},v^i_{j+1}) \in~E$. The so-called \emph{preferred path} of a worker is denoted $P^*=\langle v^*_{1},v^*_{2},...,v^*_{n} \rangle$, where $v^*_{1}$ represents the user's starting location $s$ and $v^*_{n}$ is the destination $d$. 



A worker $w$ is as individual who is willing to perform tasks in exchange for rewards while traveling along his/her preferred path $P^*$.  (While this is just a practical assumption, nothing in this work would prevent to consider the worker to be a device, e.g., a robot or an autonomous vehicle.)

We assume that all tasks are located on an edge of the network.  If a given task $t$ is not placed on an existing vertex $v \in V$ we, without loss of generality, replace that edge, say $(v_j, v_l)$, in $G$ with two new edges $(v_j, t)$ and $(t, v_l)$. Note that this implies that some of the vertices in a worker's path are now tasks rather than actual road intersections or the like.
Now that some of the vertices in $G$ are actually tasks with a positive reward associated to it, we further assume that every vertex $v$ has a reward $r(v)$ where $r(v) = 0$ if $v$ does not represent a task or $r(v) > 0$, otherwise.

The formal definitions for travel and detour costs and reward of a path are provided next.

\begin{definition}[Travel Cost]
Given a path $P^i= \langle  v^i_{1},v^i_{2},...,v^i_{n} \rangle $ in $G$, its travel cost is given by the sum of the costs of the edges in it, i.e.,
\[
TC(P^i)=\sum_{j=1}^{n-1}c(v^i_{j},v^i_{j+1}).
\]
\end{definition}


\begin{definition}[Detour Cost]
Given a  path $P^i= \langle v^i_{1},v^i_{2},...,v^i_{n} \rangle $ and the preferred path $P^*$, the detour cost of $P^i$ is defined as the sum of the costs of the edges in $P^i$ that do not belong to $P^*$. That is: 
\[DC(P^i,P^*)=\sum_{j=1}^{n-1}d(v^i_{j},v^i_{j+1},P^*),
\] 
where $d(v^i_{j},v^i_{j+1},P^*)= c(v^i_{j},v^i_{j+1})$ if $(v^i_{j},v^i_{j+1}) \not\subset P^*$ 
or null otherwise.
\end{definition}

\begin{definition}[Reward of a path]
Given a path $P^i= \langle  v^i_{1},v^i_{2},...,v^i_{n} \rangle $ in $G$, its total reward is given by the sum of the rewards of the vertices in it (recall that vertices which are not tasks have a null reward associated to them), i.e.,
\[
R(P^i)=\sum_{v_i \in P^i} r(v_i)
\]
\end{definition}


Finally, we  assume that a worker is willing to deviate from $P^*$ as long as the total travel cost of the new path, i.e., including the detour necessary for task completion, is not larger than a given budget $b$.

As mentioned earlier, IRTS aims to provide the user with a set of good alternative paths that offer different trade-offs between detour and reward. In order to do so, we rely on the notion of skyline queries, which was first introduced in \cite{borzsony2001skyline}. Given a $d$-dimensional data set, a skyline query returns the points that are not dominated by any other point. In the context of the IRTS problem, a path $P^i$ is not dominated if there is no other path $P^j$ with smaller detour \textit{and} higher reward than $P^i$. One interesting aspect of skyline queries is that the user does not need to determine beforehand weights for detour and closeness. The skyline is a set of equally interesting solutions in the sense they are all non-dominated, for arbitrary weights. The skyline set found for the IRTS problem can be formally defined as follows.

\begin{definition}[Skyline]
Let $\mathcal{P}$ be a set of paths in a two-dimensional cost space. A path $P^i \in \mathcal{P}$ dominates another path $P^j \in \mathcal{P}$, denoted as $P^i\prec P^j$, if
\[DC(P^i, P^*) < DC(P^j, P^*)~~\wedge~~R(P^i) \geq R(P^j) \hspace{0.5cm} \vee \]
\[ \hspace{-0.75cm} DC(P^i, P^*) \leq DC(P^j, P^*)~~\wedge~~R(P^i) > R(P^j)\]
That is, $P^i$ is better in one criteria and at least as good as  $P^j$ in the other one. The set of non-dominated paths, i.e. $\{P^i \in \mathcal{P}$ $|$ $\nexists P^j \in \mathcal{P} : P^i \prec P^j\}$, denotes the skyline.
\end{definition}

The IRTS problem can now be formally defined as follows.  (For ease of reference, Table~\ref{tab:notation}  summarizes the notation used throughout this paper.)

\begin{probdef}
{\em Given a worker $w$ with his/her corresponding preferred path $P^*=\langle v^*_{1},v^*_{2},...,v^*_{n} \rangle$ and budget $b$, and a set of available tasks $T$ (embedded in some vertices of the network $G$), the IRTS problem aims at finding the set of all non-dominated paths from $v^*_{1}$ to $v^*_{n}$, that contain at least one task\footnote{This avoids returning the original preferred path as a trivial non-interesting path.} $t_i \in T$ and whose travel cost does not exceed $b$.}
\end{probdef}

\begin{small}
\begin{table}[htb]
\setlength\extrarowheight{2pt}
    \caption{Notation.}     
    \centering
    \begin{tabular}{| l | l |}
    \hline
    \multicolumn{1}{|c|}{\textit{Notation}}  & \multicolumn{1}{c|}{\textit{Meaning}}      \\ \hline
    {$P^i= \langle v^i_{1},v^i_{2},...,v^i_{n} \rangle$}& {{A path $P^i$ ($P^*$ is the preferred one)}} \\ \hline 
    {{$v^i_{j}$} }&{{The $j$-{th} vertex in $P^i$} } \\ \hline 
    $s = v^i_{1}$, $d = v^i_{n}$ & The source and destination in $P^i$ \\ \hline
    $c(v_i,v_j)$ & Cost of the edge connecting $v_i$ to $v_j$ \\ \hline
    $d(v_i,v_j,P^* )$ & Detour of the edge $(v_i, v_j)$ w.r.t. $P^*$ \\ \hline 
    {{$d_{e}(.,.)$}}&{{Euclidean distance}}\\ \hline
    {$TC(P^i)$}&{Travel cost of path $P^i$} \\ \hline
    {$DC(P^i, P^*)$}&{Detour cost of path $P^i$ w.r.t. $P^*$} \\ \hline
    {$r(v_j)$}&{Reward of a vertex (task) $v_i$} \\ \hline
    {$R(P^i)$}&{Reward of path $P^i$} \\ \hline
    {$P^i\prec P^j$}&{$P^j$ is dominated by $P^i$}\\ \hline
    \end{tabular}
   \label{tab:notation}
\end{table} 
\end{small}

In order to establish IRTS's complexity, we note that finding  IRTS' skyline set includes finding the path with the highest reward such that its travel cost is under a given budget $b$ and the detour is minimum. 
We denote this IRTS sub-problem as the Maximum Reward Minimum Detour (MRMD) problem and show that MRMD's decision version is NP-Complete; therefore it follows that MRMD's optimization version is NP-Hard, and finally we can assert that the IRTS problem
is NP-hard.

\begin{thm}
The decision problem of MRMD, i.e., to decide whether there exists a valid path with reward at least $L_r$ and detour at most $U_d \leq b$, is NP-Complete.
\vspace{-0.2cm}
\begin{proof}
We prove this theorem by a reduction from the decision version of the Traveling Salesman Problem (TSP). Given a complete graph $G(V,E)$, where the cost of an edge $(v,u) \in E$ is given by the travel cost between vertices $v, u \in V$, the TSP aims at finding the shortest possible route from a given vertex $v$ that visits every vertex in $G$ exactly once and returns to $v$.
The decision version of the TSP asks the following question: Given a graph $G(V,E)$, is there a tour from a given vertex $v \in V$ with cost at most $T_{max}$? 


In order to transform a given TSP problem into an instance of MRMD we make the following assumptions:
\begin{itemize}
\item The starting location $v$ represents the worker's starting location and all the remaining vertices represent tasks (and are referred as such in the following).  
\item The travel cost of an edge linking task $i$ to task $j$ is defined as in TSP and the reward associated with each task 
is 1. 
Additionally, the budget $b$ in MRMD is equal to $T_{max}$.
\item The worker's destination $d$ is equal to $v$ and the preferred path is given by $P^*=\langle s,d \rangle$.
\item $L_r$ is given by $|V|-1$, i.e., the sum of the rewards of all tasks. 
\end{itemize}

The decision problem of the constructed MRMD instance is: Given a worker and the set of tasks $T$, can we find a path that includes all the tasks and whose detour is at most $b = T_{max}$? We now show that TSP has a Yes-instance if and only if MRMD has a Yes-instance. A solution to TSP visits every vertex with cost at most $T_{max}$. This means that all tasks can be completed, maximizing the reward, with cost at most $T_{max}$. We note that since the only edge in the preferred path is from $v$ to itself, the detour cost of any path is equal to its travel cost and, thus, it is up to $T_{max}$. On the other hand, if the MRMD problem has a Yes-instance, then it completes all the tasks and the detour cost is no greater than $b = T_{max}$. Therefore, the corresponding path is a TSP route with cost no more than $T_{max}$. This completes the proof.
\end{proof}
\end{thm}


\begin{corollary}
IRTS is NP-Hard
\vspace{-0.2cm}
\begin{proof}
Since solving the MRMD is required to solve IRTS and MRMD is NP-Hard, then it follows that IRTS is also NP-Hard.
\end{proof}
\end{corollary}



\section{Proposed Approaches}
\label{sec:solutions}
A straightforward approach to solve the IRTS problem is to find all the possible paths from $s$ to $d$ and add the non-dominated ones to the skyline set. However, this approach is not feasible even for very small instances. Therefore, we first propose an exact approach that finds the set of all non-dominated paths by checking a number of provably safe pruning conditions\footnote{A pruning condition is said to be safe if it does not prune any path that is part of the skyline set or that leads to an unexplored non-dominated path.} in order to shrink the search space. Thereafter, since the exact approach does not scale to larger instances due to the NP-hardness of the IRTS problem, we propose some heuristics that approximate the exact skyline.

\subsection{Exact Solution}
\label{sec:exact}
In order to find the exact skyline we follow an approach based on Dijkstra's algorithm.
The similarity lies in the point that unexplored nodes are queued with the most promising ones dequeued first and expanded, but the main difference is that in IRTS vertices may need to be expanded more than once. For instance, in order to find the path $P^1= \langle s, v_1, t_2, v_1, v_2, d \rangle$ in Figure~\ref{fig:irts}, 
vertex $v_1$ needs to be expanded twice.



We assume that paths from $s$ are maintained in a queue $Q$ and are expanded in increasing order of detour. Given a dequeued path $P^i$, extending it with a vertex $v$ does not lead to non-dominated paths, meaning it can be safely pruned, if one of the following conditions is satisfied:
\begin{enumerate}
\item[(P1)] $P^i$ does not contain any tasks and $v$ already belongs to $P^i$.
\item[(P2)] $P^i$ contains at least one task and $v$ already appears in $P^i$ after the last visited task $t_j \in P^i$.
\end{enumerate}


For instance, consider path $P=\{s, v_1\}$ in Figure~\ref{fig:irts}. 
Although $s$ is a neighbor of $v_1$, the path $P'=\{s, v_1, s\}$ can be pruned, according to (P1), since returning to $s$ would just increase the cost of the path without visiting any tasks. 
Let us now consider path $P= \langle s, v_1, t_2, v_1, v_2 \rangle$. Extending it with $v_1$, which would be feasible, does not help since $v_1$ has already been visited after task $t_2$, i.e., there can be no gain from returning to $v_1$.  Therefore, $P= \langle s, v_1, t_2, v_1, v_2, v_1 \rangle$ can be pruned according to (P2).
On the other hand, let us now consider path $P= \langle s, v_1, t_2 \rangle$ and feasibly extending it with $v_1$. Despite the fact that $v_1$ already belongs to $P$, it was visited before task $t_2$, i.e., the detour $\langle v_1, t_2, v_1 \rangle$ yielded a reward gain, therefore (P2) does not apply to path $P'= \langle s, v_1, t_2, v_1 \rangle$ and, thus, it can not be pruned. 

In the following we prove that both (P1) and (P2) are safe.

\begin{lemma}
Pruning condition (P1) is safe.
\begin{proof}
Let $P^i$ be a path from $s$ to a vertex $v$ that does not contain any tasks and visits $v$ only once. Moreover, assume that $P^j$  is an extension of $P^i$ that does not contain any tasks and visits $v$ twice, where $v$ is the last vertex visited by $P^j$.
By contradiction, let us assume that extending $P^j$ with a path $P^{vd}$ from $v$ to $d$ leads to a non-dominated path $P^{j'}$.
Similarly, assume that $P^{i'}$ is an extension of $P^i$ with $P^{vd}$, i.e., $P^{i'}$ is the concatenation of $P^i$ and $P^{vd}$. Since, $DC(P^i) \leq DC(P^j)$, then $DC(P^{i'}) \leq DC(P^{j'})$ also holds. Moreover, since both paths $P^{i'}$ and $P^{j'}$ visit the same set of tasks, i.e., the ones on path $P^{vd}$, $P^{i'}$ dominates $P^{j'}$.
This contradicts the assumption that $P^{j'}$ is a non-dominated path. Therefore, $P^j$ does not lead to non-dominated paths and can be safely pruned.
\end{proof}
\end{lemma}

\begin{lemma}
Pruning condition (P2) is safe.
\begin{proof}
Let $P^i$ be a path from $s$ to a vertex $v$ that contains at least one task and visits $v$ only once after the last visited task $t$. Moreover, assume that $P^j$  is an extension of $P^i$ that does not contain any new tasks and visits $v$ twice after visiting $t$, where $v$ is the last vertex visited by $P^j$. By contradiction, let us assume that extending $P^j$ with a path $P^{vd}$ from $v$ to $d$ leads to a non-dominated path $P^{j'}$. Similarly, assume that $P^{i'}$ is an extension of $P^i$ with $P^{vd}$. Since, $DC(P^i) \leq DC(P^j)$, then $DC(P^{i'}) \leq DC(P^{j'})$ also holds. Moreover, since both paths $P^{i'}$ and $P^{j'}$ visit the same set of tasks, i.e., the ones on path $P^i$ and on path $P^{vd}$, $P^{i'}$ dominates $P^{j'}$.
This contradicts the assumption that $P^{j'}$ is a non-dominated path. Therefore, $P^j$ does not lead to non-dominated paths and can be safely pruned.
\end{proof}
\end{lemma}

We note that by considering only the path that yields the minimum detour for a given set of tasks, we could ignore paths that are part of the skyline set. For instance, still considering Figure~\ref{fig:irts},
the path that includes tasks $t_2$ and $t_3$ and yields the minimum detour is $P^5= \langle s, v_1, t_2, v_1, v_2, d, t_3, d \rangle$ with detour 8. However, its total cost is 23, which exceeds the given budget $b=21$. On the other hand, path $P^2= \langle s, v_1, t_2, v_4, t_3, d \rangle$, with detour 14 and travel cost 19, does not exceed the budget and, thus, must be part of the skyline set, even though its detour is greater than that of path $P^5$ for the same set of tasks performed. 
This shows that it is not correct to prune paths just because another path including the same tasks but with a smaller detour has been found before.
Nonetheless, there are cases described by the following condition where those paths can be correctly pruned.

\begin{enumerate}
\item[(P3)] Let $P^i$ be a path dequeued from $Q$ and assume that the last vertex visited in $P^i$ is a task $t_i^n$. $P^i$ can be pruned if another path $P^j$ containing the set of tasks in $P^i$, say, $\{ t_i^1, t_i^2, \ldots, t_i^n \}$, in this particular order, has already been found with smaller or equal travel cost.
\end{enumerate}


For instance, consider path $P^i = \langle s, t_1, v_3, v_1, t_2 \rangle$ (from Figure~\ref{fig:irts}) with detour 13 and cost 13. Since paths are removed from $Q$ in increasing order of detour, the path $P^j = \langle s, t_1, s, v_1, t_2 \rangle$ with detour 8 and cost 13 has been previously found. As both $P^i$ and $P^j$ include tasks $t_1$ and $t_2$, in this order, $P^i$ can be pruned because its travel cost is the same of $P^j$ and it yields a greater detour than $P^j$.

Next, we prove that (P3) also safely prune paths.
\begin{lemma}
Pruning condition (P3) is safe.
\begin{proof}
Let $P^i$ and $P^j$ be two paths from $s$ to a task $t_k$ that visit the same set of tasks $T^i$ in the same order, where $t_k$ is the last visited task.
Additionaly, assume that $P^i$ is dequeud from $Q$ before $P^j$ and that $TC(P^j)>TC(P^i)$.
By contradiction, let us assume that extending $P^j$ with a path $P^{td}$ from $t_k$ to $d$ leads to a non-dominated path $P^{j'}$.
Similarly, assume that $P^{i'}$ is an extension of $P^i$ with $P^{td}$.
Since by assumption $P^{j'}$ is a non-dominated path, then $TC(P^{j'}) \leq b$. As $TC(P^j)>TC(P^i), TC(P^{i'}) <= b$ also holds. Therefore, $P^{i'}$ can be found within the budget.
Since $P^i$ is dequeud first, $DC(P^j) \geq DC(P^i)$. Moreover, since both paths $P^{i'}$ and $P^{j'}$ visit the same set of tasks, i.e., $T^i$ and the ones on path $P^{td}$, $P^{i'}$ dominates $P^{j'}$.
This contradicts the assumption that $P^{j'}$ is a non-dominated path.
Therefore, $P^{j}$ does not lead to non-dominated paths and can be safely pruned.
\end{proof}
\end{lemma}

Lastly, a path $P^i$ can also be pruned if its current cost plus the Euclidean distance from its last vertex $v_n^i$ to the destination $d$ exceeds the budget. In other words, if in the the best case, given that the Euclidean distance  is a lower bound to the road network distance, the total cost of a path to $d$ including the vertices of $P^i$ exceeds $b$, then that path can be safely pruned. This can be stated as follows:
\begin{enumerate}
\item[(P4)] A path $P^i$ whose last vertex is $v_n^i$ can be pruned if ($TC(P^i) + d_e(v_n^i,d)) > b.$
\end{enumerate}

\begin{lemma}
Pruning condition (P4) is safe.
\begin{proof}
Let us assume that the worker travels along a path $P^i$ from $s$ and whose last vertex is $v_n^i$. Since the worker is moving towards $d$, the remaining cost for completing his/her path is at least the travel cost of the shortest path connecting $v_n^i$ and $d$. As the Euclidean distance from $v_n^i$ to $d$ is a lower bound to their actual network distance, $TC(P^i) + d_e(v_n^i,d)$ is also a lower bound to the travel cost of a path from $s$ to $d$ that includes $P^i$. Therefore, if such lower bound is greater than the given budget $b$, the actual network distance is also greater than $b$ and thus $P^i$ can be safely pruned.
\end{proof}
\end{lemma}

Algorithm~\ref{alg} shows the pseudo-code of the exact approach. The first path added to the queue $Q$ contains only vertex $s$ (line 3). At each step, the path $P$ with minimum detour is dequeued from $Q$. If the last vertex $v$ of $P$ is the destination (line 10), we check whether $P$ is dominated w.r.t. to the paths in the skyline set $S$. A naive approach for checking whether a path is dominated is to perform a linear search in $S$. However, since paths are expanded in increasing order of detour, the detour of $P$ is greater or equal to the detour of all previously found paths. This means that it suffices to compare $P$ with the last path $P^j$ added to $S$. If $P$ has a higher reward than $P^j$, it is non-dominated and, thus, it is added to $S$. Moreover, if $P$ is non-dominated, it dominates $P^j$ if both paths yield the same detour. If that is the case, $P^j$ is removed from $S$ (line 12).

If the last vertex $v$ of $P$ is a task (line 13), we check whether the tasks of $P$, $T_p$, have already been found, in the same order they appear in $T_p$, with smaller cost. If so, $P$ is pruned (according to (P3)). Otherwise, $T_p$ is added to set of visited tasks with its corresponding cost (line 16). Next, if $P$ is not pruned, we expand it. For each vertex $u$ neighbor of $v$ we check whether it has already been visited in $P$ after the last task $t$ of $P$.  If not, a new path $P^u$ is created. Otherwise, $P^u$ is pruned according to (P2). Note that if $P$ does not contain any tasks, we assume that $t=s$ and, thus, we check whether $v$ has already been visited after $s$ (by applying (P1)). If the lower bound to the cost of a path to $d$ including the vertices of $P^u$ exceeds $b$, $P^u$ is pruned (line 23). $P^u$ is added to $Q$, otherwise. The algorithm stops when $Q$ is empty or when the detour cost of the removed path $P$ exceeds $b$ (line 7). Since paths are explored in increasing order of detour, no non-examined paths yield a detour smaller than $b$.

\begin{algorithm}[htb]
\small
\setstretch{0.8}
\caption{Exact IRTS Skyline}
\label{alg}
\DontPrintSemicolon
\KwIn{Preferred path $P^*=\langle v^*_{1},v^*_{2},...,v^*_{n} \rangle$, set of tasks $T=\lbrace t_{1},...,t_{m} \rbrace $, and budget $b$}
\KwOut{Skyline $S$ containing non-dominated paths w.r.t. detour and reward}
$S \gets \emptyset$\\
$P \gets \langle s \rangle$ \\
$Q.insert(P)$\;
$visited \gets \emptyset$\;
\While{$Q \neq \emptyset$}{
	$P \gets Q.pop()$\;
    \If{$DC(P) > b$}{
    	\Return $S$
    }
    $v \gets $ last vertex of $P$\;
    \If{$v=d \And P$ is not dominated}{
        	add $P$ to $S$\\
            remove any path dominated by $P$ from $S$\\
    }
    \If{$v$ is a task}{
    	$T_p \gets $ tasks visited in $P$\;
    	\If{$T_p$ has not been visited with smaller cost than $TC(P)$}{
        	$visited.add(T_p, TC(P))$
        }
        \Else{
        	$\Continue$\;
        }
    }
    $t \gets $ last task visited in $P$\;
   \For{all $(v,u) \in E$}{
    \If{$u$ was not visited after $t$ in $P$}		{
   		$P^u \gets $ extend $P$ with $u$\;
        \If{$TC(P^u) + d_e(u,d) \leq b$}{
        	$Q.insert(P^u)$\;
        }

    }
   }
   
}
\Return $S$
\end{algorithm} 

\subsection{Heuristic Solutions}
Due to the hardness of the IRTS problem, the exact approach does not scale to any but reasonably small sized instances. Therefore, we developed a few heuristics that approximate the exact skyline by prioritizing the path that yields the minimum detour for a given sequence of tasks. 

All proposed heuristics are based on a graph of tasks $TG(V^\prime, E^\prime)$, where the set of vertices $V'$ includes $s$, $d$ and a subset of ``feasible'' tasks in $T$. An edge $e \in E^\prime$ connecting two vertices $v$ and $u$ in $TG$ represents the path in $G$ between $v$ and $u$ that yields the minimum detour and is associated with its corresponding detour w.r.t. the preferred path $d(v,u, P^*)$ and travel cost $c(v,u)$. The set of vertices $V^\prime$ and edges $E^\prime$ are built as follows. First $s$ and $d$ are added to $V^\prime$. Next, we find all the paths $P^i$ from $s$, in increasing order of detour, to any task $t_i \in T$ such that $TC(P^i) \leq b$. Then, for any such $t_i$, we add $t_i$ to $V^\prime$ and create an edge $(s, t_i)$. Next, for each $t_i$ we repeat this process and create an edge between $t_i$ and any other task $t_j$ that can reached from $t_i$ with cost at most $b - c(s, t_i)$. Lastly, we connect every task in $V^\prime$ to $d$.
Figure~\ref{fig:graph} illustrates the task graph $TG$ obtained from $G$ in Figure~\ref{fig:irts}. For instance, the path in $G$ from $s$ to $t_2$ that yields the minimum detour has a total detour cost of 2 and travel cost equal to 7. Similarly, the minimum detour from $t_2$ to $d$ is 2 and the travel cost of the path that yields that detour is 12.

\begin{small}
\begin{figure}[h!]
	\centering
    \begin{tikzpicture}[scale=0.6]
    \tikzset{vertex/.style = {shape=circle,draw,minimum size=1em}}
	\tikzset{edge/.style = {->,> = latex'}}
    \tikzset{edgeU/.style = {<->,> = latex'}}
    
    \node[vertex] (s) at  (0,0) {$s$};
	\node[vertex] (t1) at  (-3,-2.5) {$t_1$};
    \node at (-3.75,-2) {\$3};
    \node[vertex] (t2) at  (0,-2.5) {$t_2$};
     \node at (0.55,-1.75) {\$4};
    \node[vertex] (t3) at  (3,-2.5) {$t_3$};
    \node at (3.75,-2) {\$5};
    \node[vertex] (d) at  (0,-5) {$d$};
	\draw[edge] (s) -- (t1) node[pos=.5, above, above, rotate=35] {3, 3};
    \draw[edge] (s) -- (t2) node[pos=.5, below, left] {2, 7};
	\draw[edge] (s) -- (t3) node[pos=.5, above, above, rotate=-35] {2, 17} ;
    
    \draw[edge] (t1) -- (d) node[pos=.5, below, below, rotate=-40] {3, 18};
    \draw[edge] (t2) -- (d) node[pos=.4, below, left] {2, 12};
	\draw[edge] (t3) -- (d) node[pos=.5, below, below, rotate=35] {2, 2} ;
    
    \draw[edgeU] (t1) -- (t2) node[pos=.5, above, above] {5, 10};
    
    \draw[edge] (t2) -- (t3) node[pos=.55, above, above] {4, 14};
	    

    \end{tikzpicture}
    \caption{Graph $TG$ built for the example shown in Figure~\ref{fig:irts}. The pair of values on each edge represent the corresponding detour and travel costs, respectively.}
    \label{fig:graph}
\end{figure}
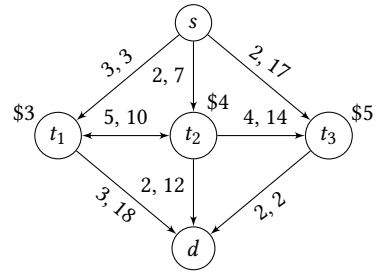
\end{small}

In order to build $TG$ we execute Dijkstra's algorithm up to $|T| + 1$ times in $G$, i.e., once for $s$ and at most once for each task in $T$. Therefore, the complexity of building $TG$ in the worst case is $O(|T|\times (|E| + |V|\times log |V|))$.

\subsubsection{Detour Oriented Heuristic (DOH)} 
Our first heuristic aims at finding all non-dominated paths in $TG$ in increasing order of detour. 
Paths are expanded from $s$ and pruned if their travel cost exceed the given budget.

Algorithm~\ref{alg:h1} shows the pseudo-code of this heuristic approach. First, the task graph $TG$ is built from $G$. Then, a path containing only vertex $s$ is added to the queue $Q$ (line 4). At each step, the path $P$ yielding the minimum detour is removed from $Q$ (line 6). If the last vertex $v$ in $P$ is $d$, we check whether it is non-dominated (as described in Section~\ref{sec:exact}) and, if so, it is added to the skyline set $S$. Note that a path that ends with $d$ does not need to be further expanded.
If $v$ is a task, for each neighbor $u$ of $v$ in $TG$ we create a new path $P^u$ that includes $u$ (line 17). If the travel cost of $P^u$ plus the cost from $u$ to $d$ does not exceed $b$, we insert $P^u$ into $Q$ (lines 18 and 19). Note that $TC(P^u) + c(u,d)$ represents the cost of traveling directly to $d$ from $P^u$ and, thus, it is a lower bound to the cost of paths to $d$ containing the tasks of $P^u$. Including a new task on the way from $u$ to $d$ would just increase the overall cost of the path which, consequently, would also have a greater travel cost than the budget.  The algorithm stops when the queue is empty or when the detour cost of the next path removed from the queue exceeds $b$ (line 7).

\begin{algorithm}[htb]
\small
\setstretch{0.8}
\caption{Detour Oriented Heuristic (DOH)}
\label{alg:h1}
\DontPrintSemicolon
\KwIn{Graph $G$, preferred path $P^*=\langle v^*_{1},v^*_{2},...,v^*_{n} \rangle$, set of tasks $T=\lbrace t_{1},...,t_{m} \rbrace $, and budget $b$}
\KwOut{Skyline $S$ containing non-dominated paths w.r.t. detour and reward}
$TG(V', E') \gets $ build task graph from $G$\;
$S \gets \emptyset$\\
$P \gets \langle s \rangle$ \\
$Q.insert(P)$\;
\While{$Q \neq \emptyset$}{
	$P \gets Q.pop()$\;
    \If{$DC(P) > b$}{
    	\Return $S$
    }
    $v \gets $ last vertex of $P$\;
    \If{$v=d$}{
            \If{$P$ is not dominated}{
        		add $P$ to $S$\\
                remove any path dominated by $P$ from $S$\\
    		}
            $\Continue$\;
    }
   \For{all $(v,u) \in E^\prime$}{
    \If{$u$ does not belong to $P$}		{
   		$P^u \gets $ extend $P$ with $u$\;
        \If{$TC(P^u) + c(u,d) \leq b$}{
        	$Q.insert(P^u)$\;
        }

    }
   }
   
}
\Return $S$
\end{algorithm}

For the graph $TG$ shown in Figure~\ref{fig:graph}, DOH would first find path $P^1 = \langle s, t_2, d \rangle$ with detour 4 and travel cost 19. Since $P^1$ is not dominated, it is added to $S$. Next, path $P^2 = \langle s, t_3, d \rangle$ is found with detour 4 and travel cost 19. Since its reward is greater than that of $P^1$ and both paths yield the same detour, $P^2$ is added to $S$ and $P^1$ is removed (as it is dominated by $P^1$). Then, path $P^3 = \langle s, t_1, d \rangle$ is found with detour 6 and cost 21. Since it is dominated, it is discarded. We note that DOH does not produce exact results. Particularly, in this example, it would not find path $P = \langle s, t_2, t_3, d \rangle$, which is part of the exact skyline. The travel cost of $P$ in $TG$ is equal to 23, which exceeds the budget $b=21$, meanwhile it would be possible to find a path including $t_2$ and $t_3$ in the original graph $G$ that would not exceed the budget.

In the worst case, DOH generates all possible permutations of tasks of size 1 to $|T|$ if each task is connected to every other task in $TG$. In this case, the maximum number of permutations is given by 
$\sum_{c=1}^{|T|} P(|T|, c)$.
For instance, for $|T|=5$, there are up to 325 permutations.

\subsubsection{$k$-NN Graph Heuristic} 
In order to avoid generating all possible permutations of tasks, the $k$-NN Graph Heuristic (kGH) limits the number of neighbors of a task $t_i$ in $TG$ to a given $k \ll |T|$, leading to $TG_{kGH}$, which is a smaller version of $TG$. More specifically, when building $TG_{kGH}$ we only connect a task $t_i$ to its $k$ closest tasks in terms of detour, which, intuitively, have a greater chance of leading to paths with shorter deviation from the preferred path $P^*$.
(We defer the discussion on what would be suitable values for $k$ to use to Section~\ref{sec:comparisons}).

In order to find the non-dominated paths in this reduced graph we follow the same procedure presented in Algorithm~$\ref{alg:h1}$ with the only difference being in line 1 where instead of $TG$ one builds $TG_{kGH}$.
We note that, when a path $P$ is dequeued from $Q$ and expanded, up to $k$ new paths of size $|P|+1$ are created. Moreover, when $|T| - |P| < k$, up to $|T| - |P|$ paths are created, since there are at most $|T| - |P|$ tasks remaining to be combined with $P$. The maximum number of paths generated by kGH is given by:
\[
|T| \times \sum_{i=1}^{|T|-k+1} (k^{i-1}) \hspace{0.15cm} + \hspace{0.15cm} |T|\times k^{|T|-k}\times \sum_{i = |T|-k+2}^{|T|} \prod_{j=1}^{|T|-i+1} (k-j)
\]

For instance, for $|T|=5$ and $k=2$, the maximum number of paths would be 115, while DOH could generate up to 325 paths.

\subsubsection{Minimum Detour Heuristic}

The Minimum Detour Heuristic (MDH) greedily expands a path $P$ with \textit{the} vertex that leads to a minimum total detour. The pseudo-code of MDH is shown in Algorithm~\ref{alg:mdh}. The first steps are similar to Algorithm~\ref{alg:h1}, but the algorithms differ from each other when expanding a path. Let $v$ be last vertex of $P$. While DOH expands $P$ with every neighbor of $v$ that does not belong to $P$, MDH only expands $P$ with a single vertex $u$, chosen from the set of neighbors. More specifically, it selects the vertex $u$ that minimizes $DC(P) + d(v,u)$ (line 3). A new path $P^u$ including $u$ is built (line 4) and if the lower bound cost to $d$ including the vertices of $P^u$ does not exceed the budget, $P^u$ is inserted into $Q$ (line 6).

We note that the MDH heuristic generates up to $|T|^2$ permutations of tasks. For each path size, in terms of number of tasks, up to $|T|$ paths are generated. Since each path may contain from 1 to $|T|$ tasks, in the worst case $|T|^2$ permutations of tasks can be examined.

\begin{algorithm}[thb]
\small
\setstretch{0.8}
\caption{Minimum Detour Heuristic (MDH)}
\label{alg:mdh}
\DontPrintSemicolon
\KwIn{Graph $G$, preferred path $P^*=\langle v^*_{1},v^*_{2},...,v^*_{n} \rangle$, set of tasks $T=\lbrace t_{1},...,t_{m} \rbrace $, and budget $b$}
\KwOut{Skyline $S$ containing non-dominated paths w.r.t. detour and reward}
	lines 1 to 14 from algorithm \ref{alg:h1}\\
    $neig \gets $ neighbors of $v$ in $TG$ that do not belong to $P$\\
    $u \gets $ vertex from $neig$ such that $DC(P) + d(v,u)$ is minimized\;
    $P^u \gets $ extend $P$ with $u$\;
        \If{$TC(P^u) + c(u,d) \leq b$}{
        	$Q.insert(P^u)$\;
        } 
\Return $S$
\end{algorithm}

\subsubsection{Maximum Reward Heuristic}
The Maximum Reward Heuristic (MRH) is similar to MDH except that, instead of expanding a path with the task that yields the minimum total detour, it selects the task with the highest reward among the neighbors of the last vertex $v$ of $P$ that are not part of $P$. As in MDH, up to $|T|^2$ permutations of tasks can be generated by MRH in the worst case.
\section{Experiments}
\label{sec:experiments}
\begin{figure}[htb]
      \begin{subfigure}[b]{0.125\textwidth}
      \centering
        \includegraphics[width=1\textwidth]{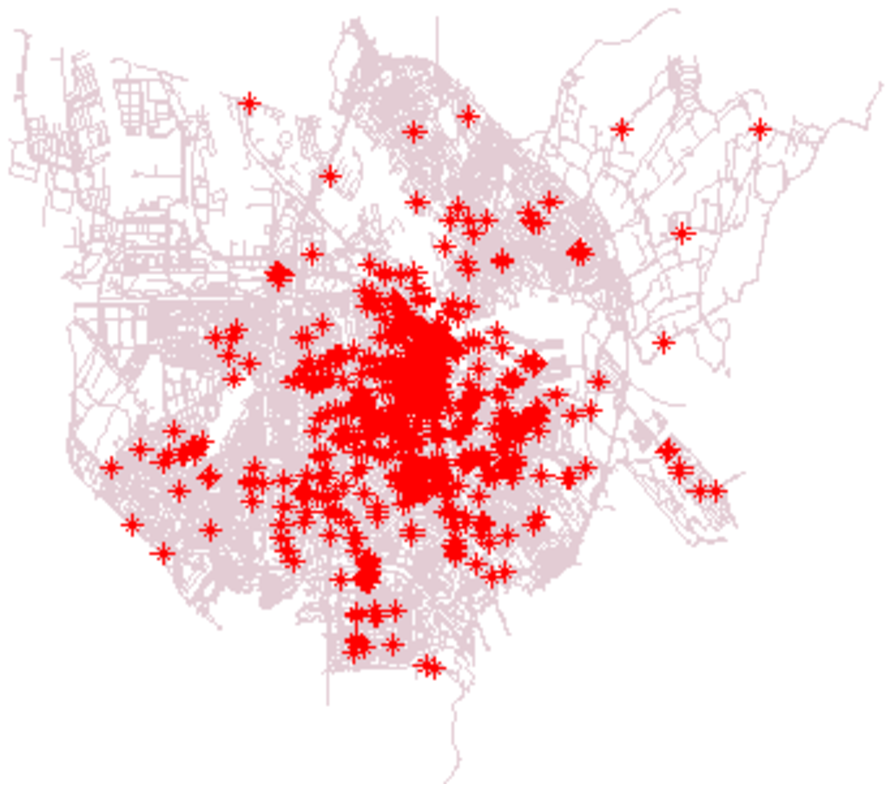}
        \caption{Amsterdam}
        \label{ams}
    \end{subfigure} 
    \quad
    \centering
    \begin{subfigure}[b]{0.125\textwidth}
          \centering
        \includegraphics[width=1\textwidth]{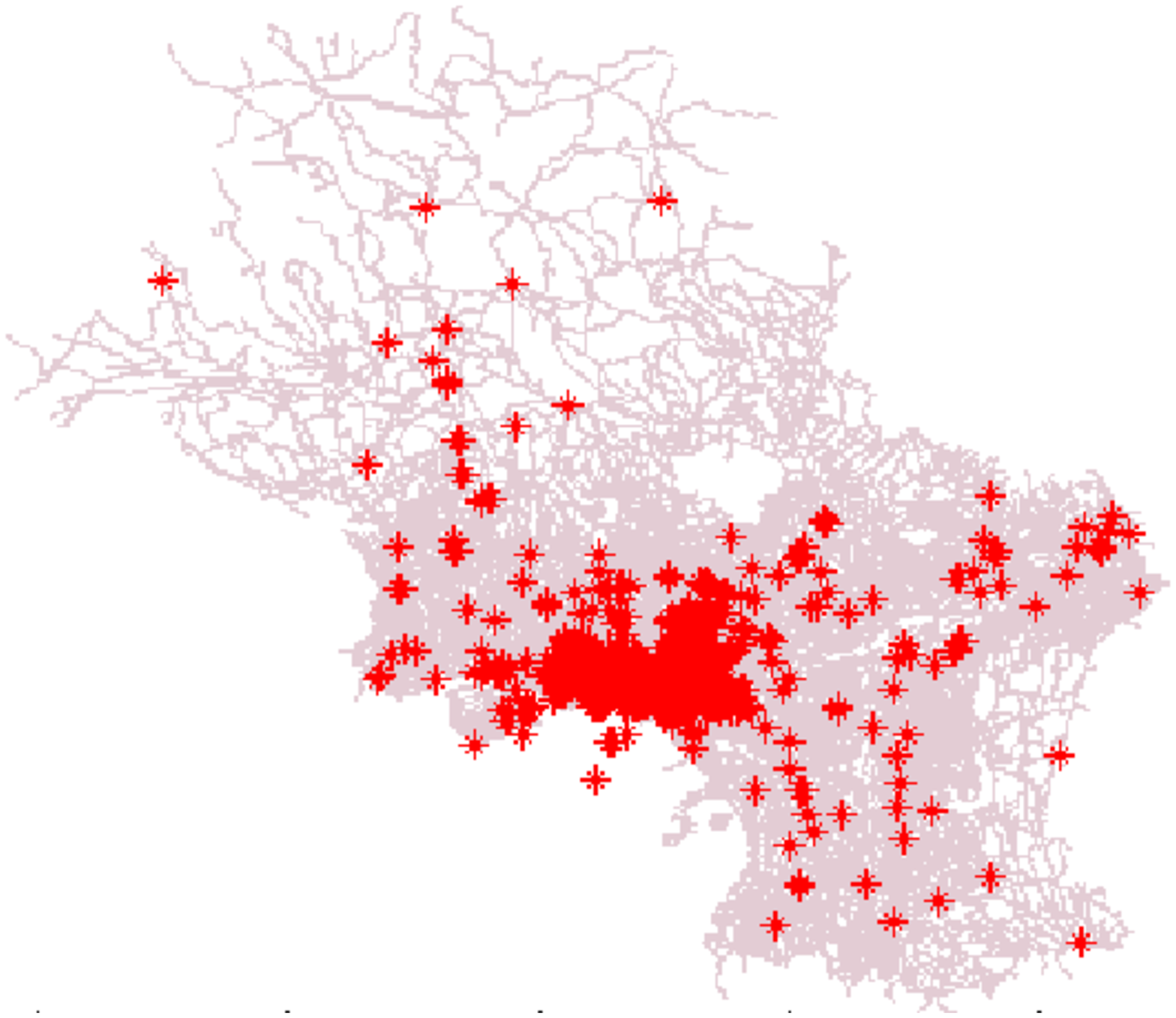}
        \caption{Oslo}
    \label{oslo}
    \end{subfigure}
    \quad
          \begin{subfigure}[b]{0.125\textwidth}
      \centering 
        \includegraphics[width=1\textwidth]{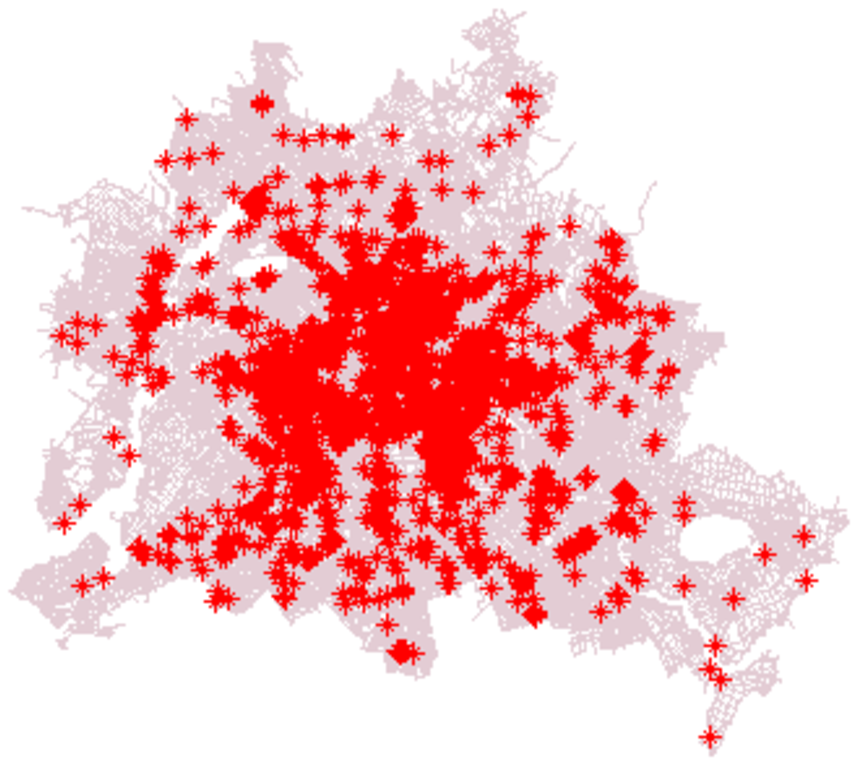}
        \caption{Berlin}
        \label{berlin}
    \end{subfigure} 
    \caption{Locations of tasks 	    
in Amsterdam, Oslo and Berlin (overlaid on those cities' road networks).}
    \label{fig:maps}
\end{figure}

We evaluated the performance of our approaches, as well as the accuracy of the approximation algorithms, varying several parameters and using real road networks. The real datasets used in our experiments reflect the road networks of Amsterdam (AMS), Oslo (OSLO) and Berlin (BER), as of March/2017 \cite{AhmadiDatasets}.
In order to have a somewhat realistic set of task locations, which could affect query processing time adversely, we use the location of eateries (restaurants and coffee shops) on those networks as locations of (pseudo) tasks. Table~\ref{Tdataset} summarizes the details of the datasets used in our experiments and Figure~\ref{fig:maps} illustrates the used road networks and the location of tasks onto them.

\begin{small}
\begin{table}[htb!]
\centering 
\caption{ Summary of the real datasets used in our experiments ({\bf bold} defines default values).}
\label{Tdataset}
\begin{tabular}{ c | c c c }
  & {Amsterdam} & {{\bf Oslo}} & {Berlin}  \\ \hline  \hline
{\#vertices} & 106,599 & {\bf 305,174} & 428,768   \\ \cline{2-4} \hline
{\#edges} & 130,090 & {\bf 330,632}  & 504,228     \\ \cline{1-4} \hline
\#tasks & 824 & {\bf 958} &  3,083
\end{tabular}
\end{table}
\end{small}

Table~\ref{tab:parameters} shows the parameters varied in our experiments, besides the real datasets. The preferred path's cost ranges from 500m to 10km. We consider that the shortest path between two locations is selected as the traveler's preferred path. It is important to stress that this is not a requirement, in fact \emph{any} path could be used, we resorted to using shortest paths only for simplicity. We assume that the travel budget is between 10\% and 50\% longer than the corresponding preferred path. 
Inspired by the experiments in \cite{deng2013maximizing}, we also varied the number of tasks, $|T|$, available to the worker between 10 and 40 tasks selected (randomly and off-line) among all tasks that can actually be completed within the given budget. (If there are less than $|T|$ feasible tasks we use only those that are feasible.)
Lastly, we also varied the distribution of rewards paid out to workers. We assumed the rewards to have either the same value or to follow a uniform (with values from 1 to 20) or exponential distribution ($\lambda=1$).

For each set of experiments, we vary the value of one parameter, and fix the other parameters to their default values. Moreover, we ran 50 cases and report the average of the results. The experiments were performed in a virtual machine with Intel(R) Xeon(R) CPU E5-2650 (8 cores @ 2.30GHz) and 16GB RAM, running Ubuntu.

\begin{small}
\begin{table}[htb]
\centering
\caption{Experimental parameters and their values ({\bf bold} defines default values).}
\label{tab:parameters}
\begin{tabular}{ l | l }
      {Parameter }&{Range}  \\ \hline \hline
	  {Cost of preferred path (km) } 
      & {0.5, 1, \textbf{2.5}, 5, 10}  \\  \hline 
      {Budget (w.r.t. $TC(P^*)$) } 
      & {110\%, \textbf{125\%}, 150\%}  \\  \hline 
      {$|T|$} 
      & {10, \textbf{20}, 40}  \\  
    \end{tabular}
\end{table}
\end{small}

\subsection{Comparison with the Exact Approach}
\label{sec:comparisons}
The exact (EXCT) approach is only practical for small instances. As we shall see shortly, query processing time increases more sharply with path length and, given that EXCT is an expensive solution, we only report its results for preferred paths that are up to 500 meters long, with the other parameters fixed to their default values. In order to compare the non-exact approaches w.r.t EXCT, we evaluated their \emph{recall}, i.e., the percentage of non-dominated solutions (the skyline set) found by them, as well as their \emph{precision}, i.e., percentage of non-dominated solutions in their result sets.
Ideally, one wants both of those to be as close as possible to 100\%. 

\begin{figure}[htb]
      \begin{subfigure}[b]{0.22\textwidth}
      \centering
        \includegraphics[width=1\textwidth]{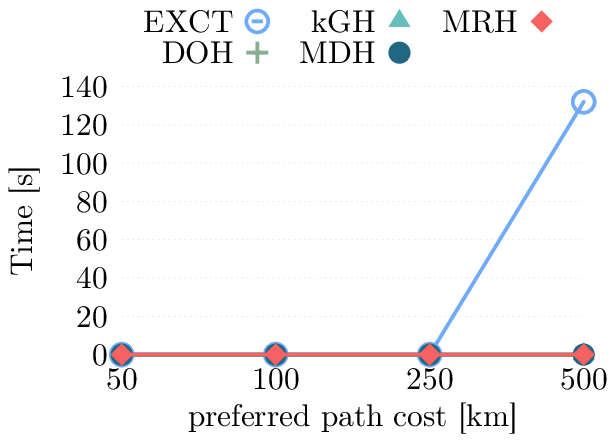}
        \caption{time}
        \label{fig:short_time}
    \end{subfigure} 
    
    \begin{subfigure}[b]{0.22\textwidth}
      \centering
        \includegraphics[width=1\textwidth]{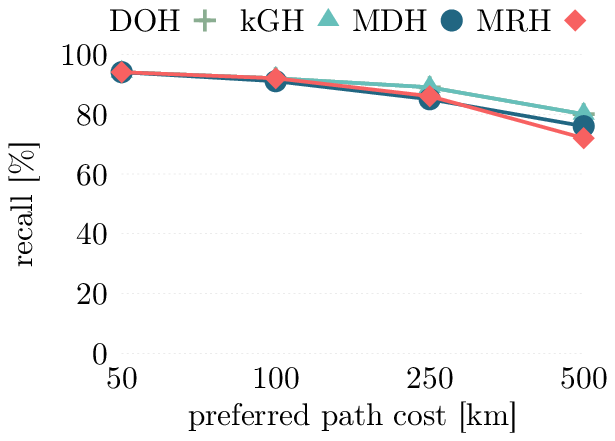}
        \caption{recall}
        \label{fig:short_missing}
    \end{subfigure} 
    \quad
    \begin{subfigure}[b]{0.22\textwidth}
      \centering
        \includegraphics[width=1\textwidth]{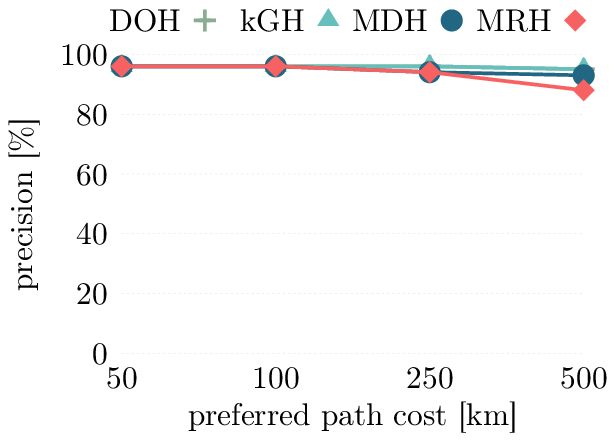}
        \caption{precision}
        \label{fig:short_dominated}
    \end{subfigure} 
    \caption{Processing time and effectiveness w.r.t. preferred path cost.}
    \label{fig:short}
\end{figure}

As shown in Figure~\ref{fig:short_time}, EXCT's performance degrades very fast when the path cost increases. It required 132 seconds in average for 500 meters long paths, which strongly suggests that it is not a practical alternative for any non-trivially sized problem.  In sharp contrast, the heuristic solutions required in average only a few milliseconds to solve the same problems. Interestingly, we noted that the variation between the processing times for EXCT in the OSLO network is very high (between 76ms and 15min for 500 meters long paths). This can be explained by the distribution of tasks in that network. Most tasks are concentrated in the same area, as shown in Figure~\ref{oslo}. Since the starting points of preferred paths are randomly selected, some paths will be in a very dense area, in terms of number of tasks, while others will be in sparse areas. EXCT tends to prune less paths when there is a high number of tasks to choose from, which leads to a sharp increase in the processing time.

As shown in Figures~\ref{fig:short_missing} and \ref{fig:short_dominated}, and as expected, DOH produces the best results w.r.t. EXCT. It finds at least 80\% of the exact skyline, while only around up to 5\% of the results produced by it are dominated, or alternatively, at least 95\% of the solution returned is made of true-positives. Although kGH produced results similar to the ones of DOH, the difference between their results tends to increase with the preferred path length (we discuss that further shortly).

\subsection{Results for the Approximate Approaches}
Since EXCT is not feasible for non-trivially sized instance of the IRTS problem, and given the lack of alternative, in what follows we use the results produced by DOH as an ``optimistic'' ground truth given that it is the heuristic that better approximates the optimal skyline. Let $S_{EXCT}$ and  $S_{DOH}$ be the skyline sets found by EXCT and DOH, respectively. If a path $P^i$ is dominated by any path in $S_{DOH}$, it is also dominated by $S_{EXCT}$. However, if a path is non-dominated by $S_{DOH}$, it may or may not be part of $S_{EXCT}$. Therefore, the precision and recall values presented next are optimistic, i.e., an upper bound of the actual values w.r.t. the exact skyline (if they could be obtained).



Before presenting all results, recall that the kGH approach has $k$ as an input parameter.  In order to determine which value to use we ran experiments using different values of $k \in \{1, 2, 5, 10\}$, while keeping all other parameters at their default values.  We found out that the average optimistic recall varied between 55\% and 99\% 
and average prevision values were in the much tighter range between 82\% and 100\%, respectively.  Average processing time ranged between 126 ms and 176 ms, respectively.  We observed that for $k = 5$ both optimistic recall and precision were above 90\% and using $k = 10$ improved those numbers only slightly at the cost of about 10\% more processing time. Given these preliminary results, we decided to use kGH with $k = 5$ in all experiments that follow.

\subsubsection{Effect of the road network}
\begin{figure}[htb]
      \begin{subfigure}[b]{0.22\textwidth}
      \centering
        \includegraphics[width=1\textwidth]{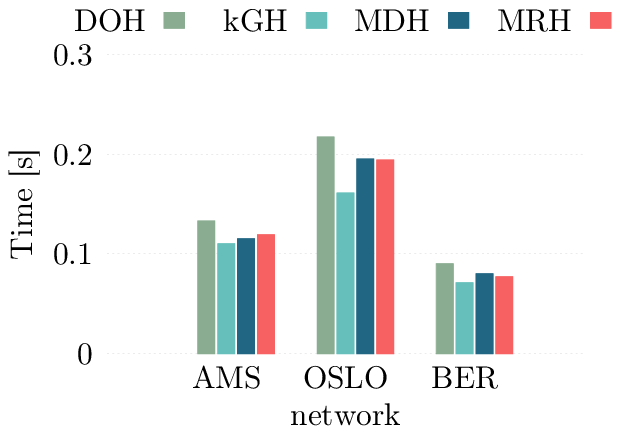}
        \caption{time}
        \label{fig:city_time}
    \end{subfigure} 
    
    \begin{subfigure}[b]{0.22\textwidth}
      \centering
        \includegraphics[width=1\textwidth]{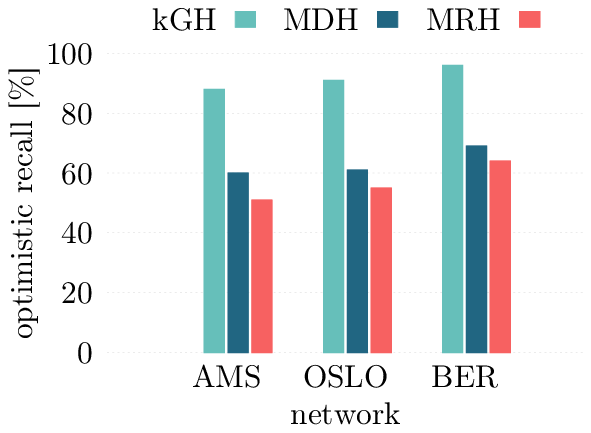}
        \caption{optimistic recall}
        \label{fig:city_missing}
    \end{subfigure} 
    \quad
    \begin{subfigure}[b]{0.22\textwidth}
      \centering
        \includegraphics[width=1\textwidth]{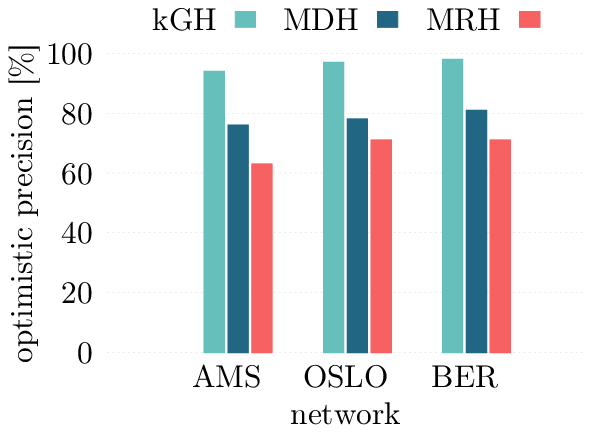}
        \caption{optimistic precision}
        \label{fig:city_dominated}
    \end{subfigure} 
    \caption{Processing time and effectiveness w.r.t. network.}
    \label{fig:network}
\end{figure}

As shown in Figure~\ref{fig:city_time}, the processing time of the heuristics is greatly affected by the networks. The more concentrated the potential tasks are around the same area, as is the case for OSLO (Figure~\ref{oslo}), the greater the probability of more tasks to be included within the budget.
This directly affects the processing time of all heuristics since more tasks will be part of the task graph.
All heuristics produced slightly worse results for the AMS network, as shown in Figures ~\ref{fig:city_missing} and \ref{fig:city_dominated}. When compared to OSLO and BER, AMS tends to include more cases where tasks that have the potential to compose a new non-dominated set are significantly far, in terms of detour cost, from previously found tasks. This, in turn, potentially increases the likelihood of these tasks being missed by the heuristics and excluded from the skyline set found by them.

\subsubsection{Effect of the cost of the preferred path}
\begin{figure}[htb]
      \begin{subfigure}[b]{0.22\textwidth}
      \centering
        \includegraphics[width=1\textwidth]{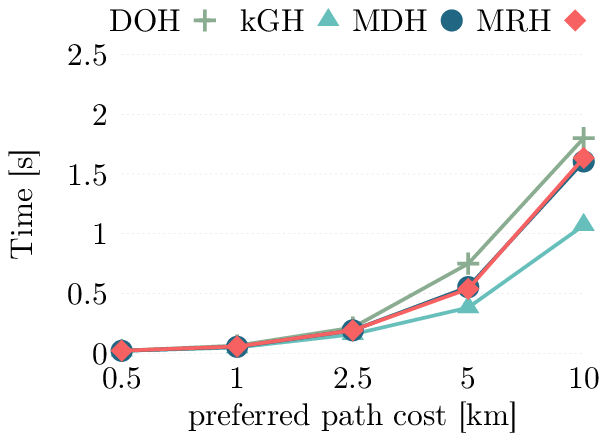}
        \caption{time}
        \label{fig:cost_time}
    \end{subfigure} 
    
    \begin{subfigure}[b]{0.22\textwidth}
      \centering
        \includegraphics[width=1\textwidth]{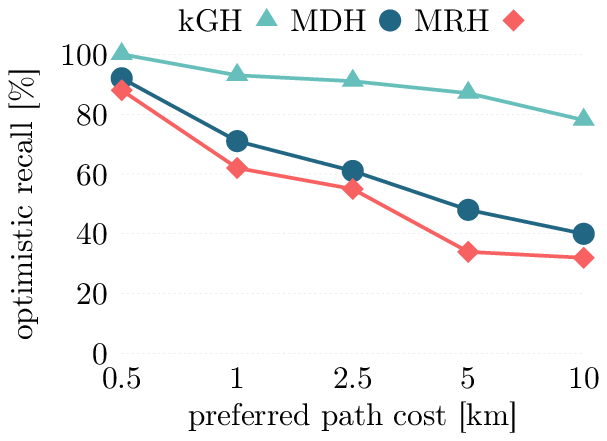}
        \caption{optimistic recall}
        \label{fig:cost_missing}
    \end{subfigure}
    \quad
    \begin{subfigure}[b]{0.22\textwidth}
      \centering
        \includegraphics[width=1\textwidth]{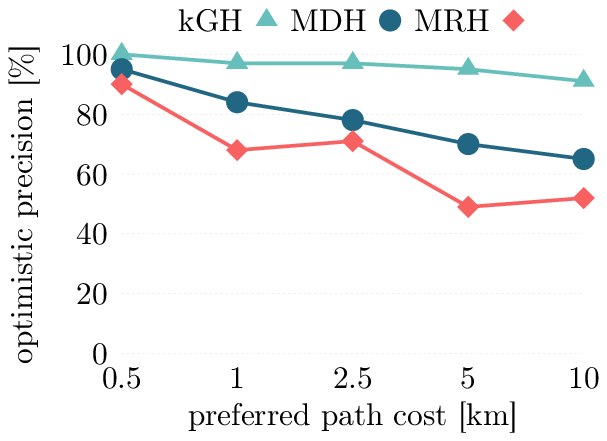}
        \caption{optimistic precision}
        \label{ams}
    \end{subfigure} 
    \caption{Processing time and effectiveness w.r.t. preferred path cost.}
    \label{fig:cost_dominated}
\end{figure}

Figure~\ref{fig:cost_time} shows that, as expected, the processing time of our heuristics increases with the path length, since more tasks will tend to be around a longer path. Moreover, a longer path also means a greater budget. This, in turn, implies that more tasks can be performed in sequence, which potentially leads to an increase in the number of times that paths will be expanded by the proposed approaches.
As in the previous experiment, kGH performed better both in terms of optimistic recall and precision than the greedy heuristics MDH and MRH, while being faster. This can be explained by the fact that, when a path $P$ is removed from the queue, up to $k$ new paths of size $|P|+1$ are created by kGH, while MDH and MRH only generate \textit{one} new path greedily. Therefore, kGH is more likely to produce better paths. Moreover, MDH and MRH expand the original task graph $TG$ where there are $O(|T|^2)$ edges between tasks. On the other hand, kGH looks for non-dominated paths in a reduced graph with $O(|T|\times k)$ edges between tasks, which explains why kGH outperforms the other two heuristics. 

\subsubsection{Effect of the budget}
\begin{figure}[htb]
      \begin{subfigure}[b]{0.22\textwidth}
      \centering
        \includegraphics[width=1\textwidth]{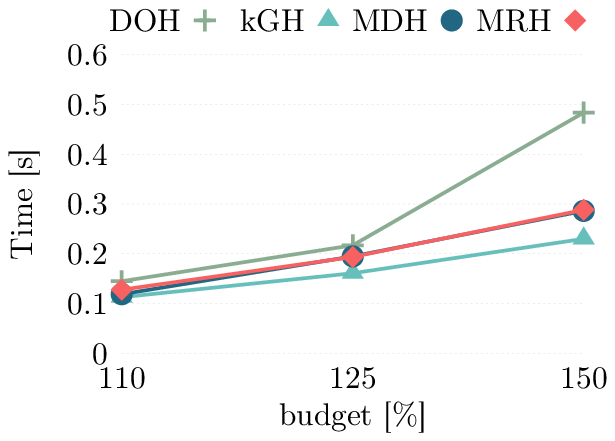}
        \caption{time}
        \label{fig:budget_time}
    \end{subfigure} 
    
    \begin{subfigure}[b]{0.22\textwidth}
      \centering
        \includegraphics[width=1\textwidth]{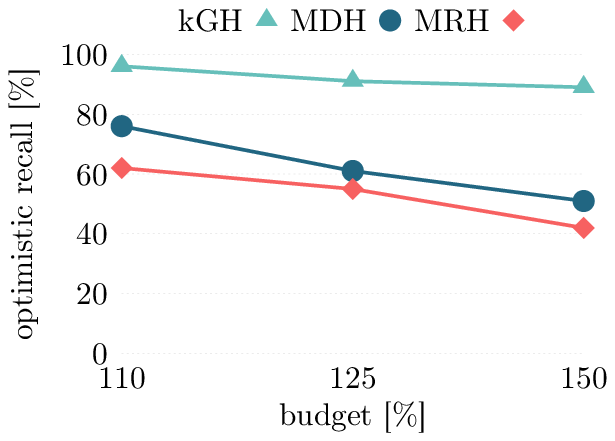}
        \caption{optimistic recall}
        \label{fig:budget_missing}
    \end{subfigure} 
    \quad
    \begin{subfigure}[b]{0.22\textwidth}
      \centering
        \includegraphics[width=1\textwidth]{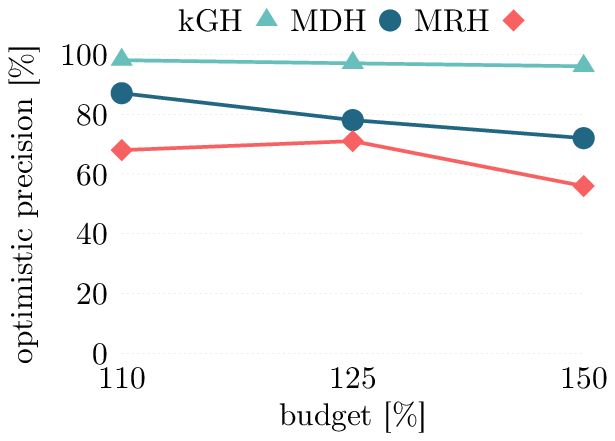}
        \caption{optimistic precision}
        \label{fig:budget_dominated}
    \end{subfigure} 
    \caption{Processing time and effectiveness w.r.t. budget.}
    \label{fig:budget}
\end{figure}

As expected and shown in Figure~\ref{fig:budget_time}, the processing time of all approaches increases with the budget, simply because there are more feasible tasks to be considered within the given budget. However, note that increasing the budget, for paths with length 2.5km, tends to include lesser new tasks than when compared to increasing the path length to 5k or 10km. Moreover, the number of tasks that can be performed in sequence is also not as high. This explains why the processing time does not increase as fast as in the experiment where the path length was varied. Also, note that, differently from the results shown in Figure~\ref{fig:cost_time}, DOH is more affected by an increasing in the budget than the other approaches. DOH will tend to check more permutations of tasks, since it strives to find all possible non-dominated paths in the task graph.
%
For the same reasons explained above, as shown in Figures~\ref{fig:budget_missing} and \ref{fig:budget_dominated}, the optimistic recall and precision of the solutions produced by the heuristic approaches are not as affected when the budget increases as when the path length is varied. A smaller increase in the number of tasks reduces the risk of choosing a task poorly.

\subsubsection{Effect of $|T|$}
\begin{figure}[htb]
      \begin{subfigure}[b]{0.22\textwidth}
      \centering
        \includegraphics[width=1\textwidth]{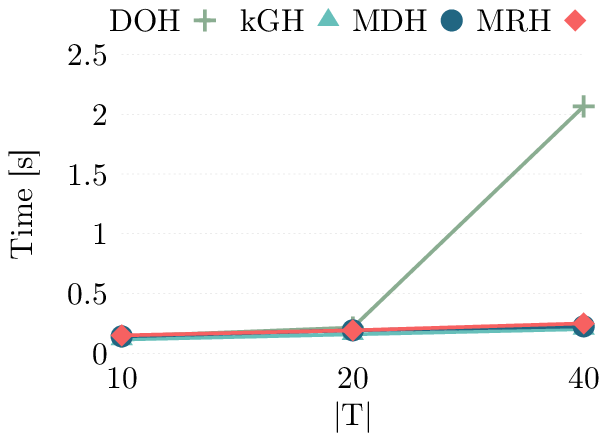}
        \caption{time}
        \label{fig:num_time}
    \end{subfigure} 
    \quad
    \begin{subfigure}[b]{0.22\textwidth}
      \centering
        \includegraphics[width=1\textwidth]{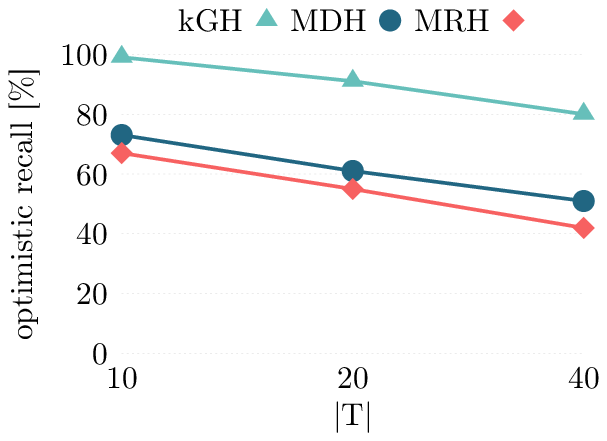}
        \caption{optimistic recall}
        \label{fig:num_missing}
    \end{subfigure} 
    
    \begin{subfigure}[b]{0.22\textwidth}
      \centering
        \includegraphics[width=1\textwidth]{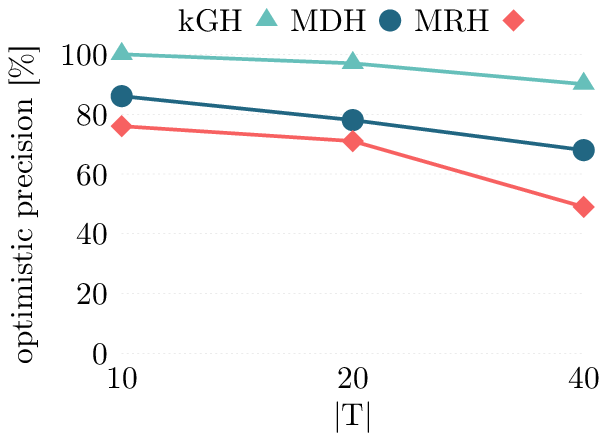}
        \caption{optimistic precision}
        \label{fig:num_dominated}
    \end{subfigure} 
    \caption{Processing time and effectiveness w.r.t.  |T|.}
    \label{fig:num}
\end{figure}

Figure~\ref{fig:num_time} shows that, while the processing time of kGH, MDH and MRH increases only slightly with the number of feasible tasks, DOH is greatly affected by this parameter. This is due to the high number of permutations of tasks that DOH may check when looking for non-dominated paths. 
As shown in Figures~\ref{fig:num_missing} and \ref{fig:num_dominated}, the optmistic recall and precision of the solutions produced by the heuristic approaches also decrease when the number of tasks increases, as also evidenced in the previous experiments. Intuitively, the likelihood of poorly choosing a task when expanding a path increases with the number of tasks.

\subsubsection{Effect of the distribution of rewards}
\begin{figure}[htb]
      \begin{subfigure}[b]{0.22\textwidth}
      \centering
        \includegraphics[width=1\textwidth]{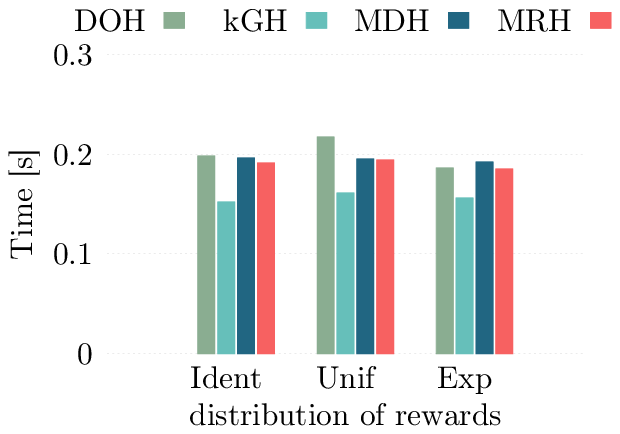}
        \caption{time}
        \label{fig:dist_time}
    \end{subfigure} 
    
    \begin{subfigure}[b]{0.22\textwidth}
      \centering
        \includegraphics[width=1\textwidth]{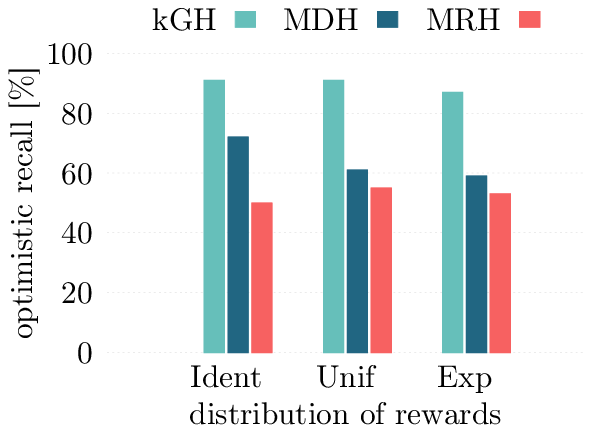}
        \caption{optimistic recall}
        \label{fig:dist_missing}
    \end{subfigure} 
    \quad
    \begin{subfigure}[b]{0.22\textwidth}
      \centering
        \includegraphics[width=1\textwidth]{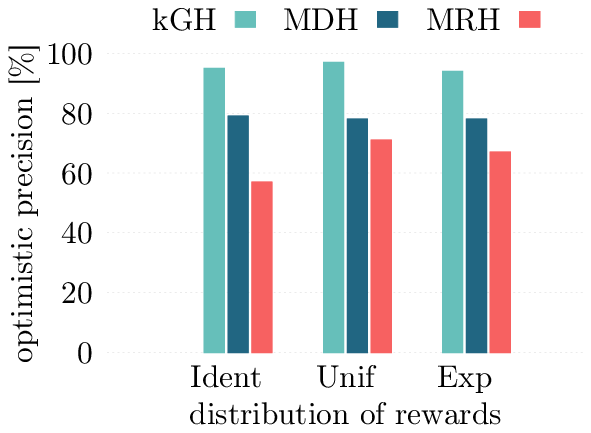}
        \caption{optimistic precision}
        \label{fig:dist_dominated}
    \end{subfigure} 
    \caption{Processing time and effectiveness w.r.t. distribution of rewards.}
    \label{fig:dist}
\end{figure}

Figure~\ref{fig:dist_time} suggests that the processing time of the heuristic approaches is not sensitive to the variation of distributions of rewards.
%
Regarding effectiveness, Figures~\ref{fig:dist_missing} and \ref{fig:dist_dominated} show that MRH is the most sensitive approach to this parameter, as expected. When all tasks have the same reward, MRH chooses any task to expand a path with, which increases the chance of picking one that leads to poor results. On the other hand, for the exponential distribution, since some tasks will have very high reward, those tasks will tend to be chosen by MRH. However, since the proximity from the chosen task to the tasks already in the path is not taken into consideration, MRH may miss tasks that are closer to the current path and, in turn, could lead to a longer sequence of tasks. For the uniform distribution, MRH will not suffer from any of these drawbacks and, thus, as evidenced in this experiment, produces better results. Moreover, Figure~\ref{fig:dist_missing} also shows that MDH presents significantly higher recall when all rewards are the same. Since MDH prioritizes the task that yields the smallest detour, and in this case all tasks have the same reward, it will tend to find tasks that are good both in terms of detour and reward. 

\subsubsection{Effect of the distribution of tasks}
\begin{figure}[b!]
      \begin{subfigure}[b]{0.22\textwidth}
      \centering
        \includegraphics[width=1\textwidth]{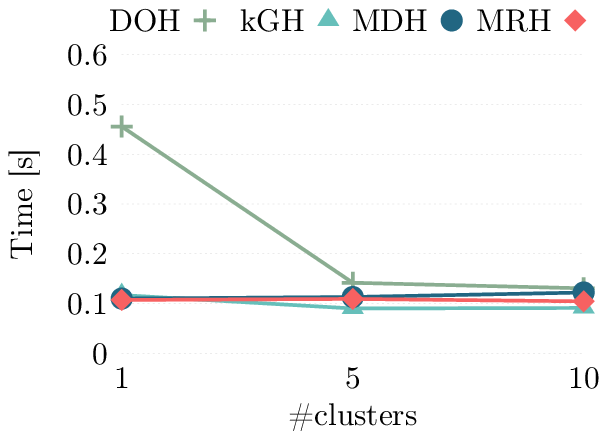}
        \caption{time}
        \label{fig:cluster_time}
    \end{subfigure} 
    
    \begin{subfigure}[b]{0.22\textwidth}
      \centering
        \includegraphics[width=1\textwidth]{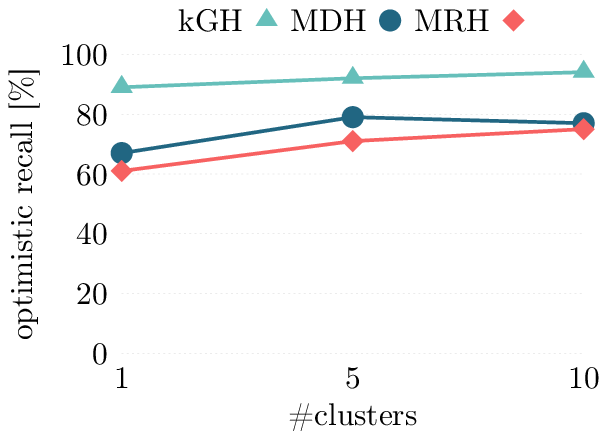}
        \caption{optimistic recall}
        \label{fig:cluster_missing}
    \end{subfigure} 
    \quad
    \begin{subfigure}[b]{0.22\textwidth}
      \centering
        \includegraphics[width=1\textwidth]{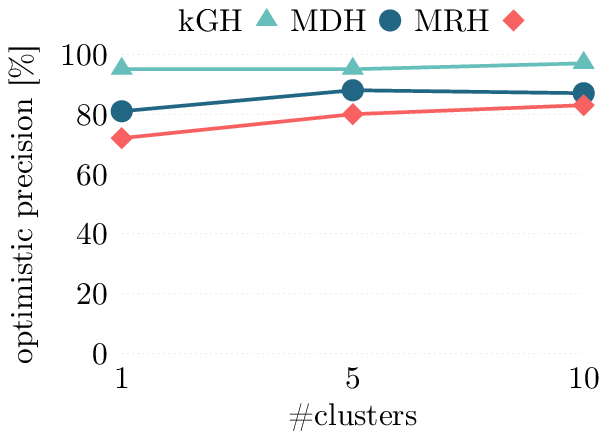}
        \caption{optimistic precision}
        \label{fig:cluster_dominated}
    \end{subfigure} 
    \caption{Processing time and effectiveness w.r.t. the number of clusters.}
    \label{fig:cluster}
\end{figure}

Even though we use the actual location of eateries as proxy for task locations, in all experiments above the $|T|$ tasks to be performed where sampled randomly among all those feasible.
In this last set of experiments we evaluated how our proposed approaches behave under a clustered distribution of tasks. 
More specifically, given a number of clusters $c$, each cluster centroid plus their $|T|/c - 1$ nearest neighbors are considered to be the available tasks.  This ensures the total number of tasks is $|T|$ while still distributing them within the $c$ clusters.
We varied the number of clusters $c$ from 1 to 10 and the results obtained are reported in Figure~\ref{fig:cluster}. As shown in Figure~\ref{fig:cluster_time}, DOH is the most affected approach in terms of processing time. The more concentrated the tasks are and the closer they are from each other, the more likely it is that more task sequences will be within the budget. Since all those task sequences are examined by DOH, the smaller the number of clusters, the longer it takes to find the skyline set.

The quality of the results produced by our heuristics tends to increase with the number of clusters, as shown in Figures~\ref{fig:cluster_missing} and \ref{fig:cluster_dominated}. The greater the number of clusters, the more scattered the points will tend to be. Consequently, less task sequences will be within the cost budget. This potentially reduces the likelihood of making a poor decision when extending a path with a new task. We also note that kGH is the least affected by this parameter since it examines up to $k$ candidate tasks when expanding a path, differently from MDH and MRH that only examine one candidate.

\section{Conclusion}
\label{sec:conclusion}

In this paper we presented the IRTS problem, a new variation of the spatial crowdsourcing problem which considers that the worker has (or is on) a preferred path and is willing to consider the trade-off between a limited detour and rewards collected by completing tasks during such detour.  We investigated this problem using a skyline approach, and given IRTS's NP-hardness, we proposed a few heuristic solutions.  Our experimental results, using real datasets at the city scale, showed that our proposed solutions, notably the one guided by the detour's length (DOH), can obtain very good solutions for IRTS instances of realistic size often in under one second, making it of practical interest.  A direction for future work is to incorporate the temporal dimension into IRTS, e.g., by considering not only travel time but also the time to complete tasks and considering an upper-limit to both travel distance as well as travel time. 

\section*{Acknowledgements}
Research partially supported by NSERC, Canada and CNPq's Science Without Borders Program, Brazil.

\bibliographystyle{ACM-Reference-Format}
\bibliography{sample-bibliography}

\end{document}